\documentstyle[psfig]{mn}

\title[VLBI Observations of faint GPS sources]{Multifrequency VLBI observations of faint gigahertz peaked spectrum sources}

\author[I. Snellen et al.]{I.A.G. Snellen$^{1,2}$, R.T. Schilizzi$^{2,3}$,
H.J van Langevelde$^{3}$, \\ 
$^{1}$Institute of Astronomy, Madingley Road, Cambridge CB3 0HA, United 
Kingdom\\
$^{2}$Leiden Observatory, P.O. Box 9513, 2300 RA, Leiden, The Netherlands \\
$^{3}$Joint Institute for VLBI in Europe, Postbus 2, 7990 AA, Dwingeloo, 
The Netherlands}

\date{}
\begin{document}
\maketitle
\begin{abstract}
We present the data and analysis of VLBI observations at 1.6, 5 and 15
GHz of a sample of faint Gigahertz Peaked Spectrum (GPS) sources
selected from the Westerbork Northern Sky Survey (WENSS).  The 5 GHz
observations involved a global array of 16 stations and yielded data
on the total sample of 47 sources.  A subsample of 26 GPS sources with
peak frequencies $\nu_p > 5$ GHz and/or peak flux densities $S_p > 125
$ mJy was observed with the VLBA at 15 GHz. A second subsample of 29 sources,
with $\nu_p <5$ GHz, was
observed at 1.6 GHz using a 14 station global VLBI array.
In this way, 44 of the 47 sources (94\%) in the sample were observed 
above and at or below their spectral peak.
Spectral decomposition allowed us to identify 3, 11, 7, and 2
objects as compact symmetric objects, compact doubles, core-jet and 
complex sources respectively. 
 However, many of the sources classified 
as compact double or core-jet sources show only two components making 
their classification rather tentative. This may explain why
the strong morphological dichotomy of GPS quasars and galaxies found for 
radio-bright GPS sources, is not as clear in this faint sample.
\end{abstract}

\section{Introduction}

Gigahertz Peaked Spectrum (GPS, e.g. O'Dea 1998) are
a class of extragalactic radio source, characterised by a convex shaped
radio spectrum peaking at about 1 GHz in frequency, and sub-galactic sizes.
Their small sizes make observations using
Very Long Baseline Interferometry (VLBI) necessary to reveal their 
radio morphologies. Early VLBI observations showed that some GPS sources 
identified with galaxies have Compact Double (CD) morphologies (Philips
and Mutel, 1982), and it was suggested that 
these were the mini-lobes of very young or alternatively old, frustrated 
objects (Philips
and Mutel, 1982; Wilkinson et al. 1984, van Breugel, Miley and Heckman, 1984).
Later, when reliable VLBI observations at higher frequencies became possible, 
it was found that some of the CD-sources had a compact flat spectrum 
component in their centres (Conway et al. 1992, Wilkinson et al. 1994).
These flat spectrum components were interpreted as the central cores,
and many CD-sources were renamed compact triples or 
Compact Symmetric Objects (CSO, Conway et al. 1992, Wilkinson et al. 1994). 
High dynamic range VLBI observations by
Dallacasa et al (1995) and Stanghellini et al.
(1997) have shown that most GPS galaxies indeed have jets leading from
the central compact core to the outer hotspots or lobes.
This is in contrast to the GPS sources identified with quasars, which tend
to have core-jet morphologies with no outer lobes (Stanghellini et al. 1997). 
Snellen et al. (1999) have shown that the redshift distributions of the 
GPS galaxies and quasars are very different, and that it is therefore
unlikely that they form a single class of object unified by orientation.
They suggest that they are separate classes of object, which just happen to 
have the same radio-spectral morphologies.

The separation velocities of the hotspots have now
been measured for a small number of GPS galaxies to be $0.2h^{-1}$c 
(Owsianik and Conway, 1998; Owsianik, Conway and Polatidis, 1998; Tschager et
al. 1999).
This makes it very likely that these are young objects of 
ages typically $\sim 10^3$ yr (assuming a constant separation velocity),
rather than old objects constrained in their growth by a dense ISM. 
These are therefore the objects of choice to study the early evolution
of extragalactic radio sources.

In the past, work has been concentrated on samples of 
the radio brightest GPS sources (eg. O'Dea et al 1991).
In order to disentangle radio power and redshift effects on the properties
of GPS sources, we constructed a sample of faint GPS sources  
from the Westerbork Northern Sky Survey (WENSS, Rengelink et al. 1997),
which in combination with other samples allows, for the first time,
the study of these objects over a large range of flux density and radio 
spectral peak frequency.
The construction of the faint sample is described in Snellen et al. (1998a); 
the optical and near-infrared imaging is described in 
Snellen et al. (1998b); and the optical spectroscopy in Snellen et al. (1999a).
This paper describes multi-frequency VLBI observations of the sample, and
the  radio-morphologies of the individual sources.
What can be learned from the faint GPS sample about radio source
evolution is discussed in the accompanying paper (Snellen et al. 2000).

\section{The Sample}

The selection of the sample has been described in detail 
in Snellen et al. (1998a), and is summarised here.
Candidate GPS sources were selected from the Westerbork Northern Sky survey, 
by means of 
 an inverted spectrum between 325 MHz and higher frequencies.
The sources are located in two regions of the survey; one with $15^h < \alpha 
< 20^h$
and $58^\circ< \delta < 75^\circ$, which is called the {\it mini-survey} region
(Rengelink et al. 1997), and the other with $4^h00^m < \alpha < 8^h30^m$ and
$58^\circ< \delta < 75^\circ$. Additional observations at 1.4, 5, 8.4 and 15
GHz were carried out with the WSRT and the VLA, yielding a sample of 47
genuine GPS sources with peak frequencies ranging from 500 MHz to more than 15
GHz, and peak flux densities ranging from $\sim30$ to $\sim900$ mJy.
This sample has been imaged in the optical and near-infrared, resulting in
an identification fraction of $\sim$ 87 \% (Snellen et al. 1998b).
Redshifts have been obtained for 40\% of the sample (Snellen et al., 1999).

\section{Observations}

Snapshot VLBI observations were made of the entire sample of 
faint GPS sources 
at 5 GHz, and of sub-samples at 15 GHz and 1.6 GHz frequency.
In order to observe the large number of sources required in a reasonable
amount of time, we observed in ``snapshot'' mode (eg. Polatidis et al. 1995, 
Henstock et al. 1995).
This entails observing a source for short periods of time at several different
hour angles. Using a VLBI array of typically more than 10 telescopes, this
provides sufficient $u,v$ coverage for reliable mapping of 
complex sources (Polatidis et al. 1995).
To maximize the $u,v$ coverage for each source we attempted to schedule three 
to four scans as widely spaced as possible within the visibility window 
during which the source could be seen by all antennas. 
Fortunately, the majority of the sources are located at  a sufficiently high 
declination ($>57^{\circ}$) that they are
circumpolar for most EVN and VLBA antennas, and therefore could be scheduled for
observation at optimal hour angles.

\begin{figure}
\psfig{figure=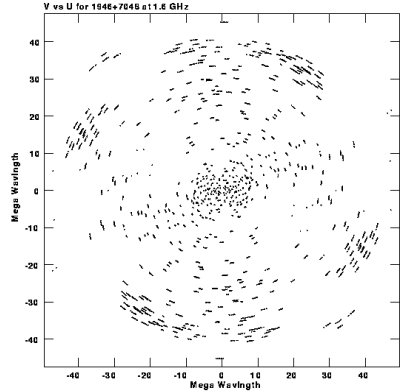,width=6cm}
\psfig{figure=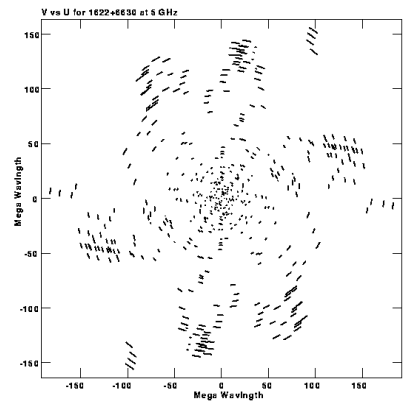,width=6cm}
\psfig{figure=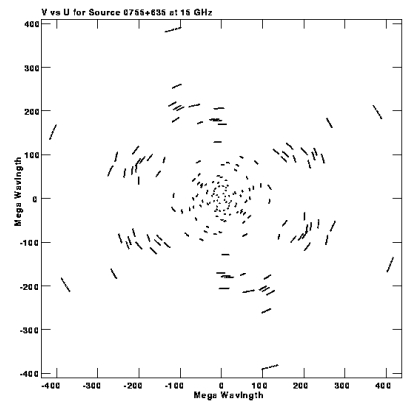,width=6cm}
\caption{\label{UV} Typical $u,v$ coverages for a source observed at 1.6 GHz 
(upper), 5 GHz (middle), 15 GHz (lower).}
\end{figure}

\begin{table*}
\centerline{
\begin{tabular}{llcccc} 
        &                      &   & &\\ 
Telescope&Location&Diam.   &SEFD$_{1.6GHz}$ & SEFD$_{5GHz}$&SEFD$_{15 GHz}$ \\
        &         &(m)&(Jy) & (Jy) & (Jy)\\ 
Cambridge  &EVN, U.K.       &32         & &136&\\
Effelsberg &EVN, Germany    &100        &19&20&\\
JB, Lovell&EVN, U.K.        &76         &44&&\\
JB, MK2   &EVN, U.K.        &25         & &320&\\
Medicina   &EVN, Italy      &32         & &296&\\
Noto       &EVN, Italy      &32        & &260&\\
Torun&EVN, Poland           &32      &230& &\\
Westerbork &EVN, Netherlands&$12\times25$&450$^*$&90&\\
VLBA\_BR&     Brewster, WA, USA&25&300&300&525\\
VLBA\_FD&   Fort Davis, TX, USA&25&300&300&525\\
VLBA\_HN&      Hancock, NH, USA&25&300&300&525\\
VLBA\_KP&    Kitt Peak, AZ, USA&25&300&300&525\\
VLBA\_LA&   Los Alamos, NM, USA&25&300&300&525\\
VLBA\_MK&    Mauna Kea, HI, USA&25&300&300&525\\
VLBA\_NL&North Liberty, IA, USA&25&300&300&525\\
VLBA\_OV&  Owens Valley, CA, USA&25&300&300&525\\
VLBA\_PT&     Pie Town, NM, USA&25&300&300&525\\
VLBA\_SC&  Saint Croix, VI, USA&25&300&300&525\\ 
\end{tabular}}
$^*$ Westerbork only observed with a single antenna at 1.6 GHz.
\caption{\label{tel} The telescopes used for the VLBI observations}
\end{table*}

\subsection{The 5 GHz Observations, Correlation and Reduction}

The 5 GHz data were obtained during a 48 hour observing session on 15 and
16 May 1995. All telescopes of the VLBA, and six
telescopes of the EVN were scheduled to participate
in this global VLBI experiment (see table \ref{tel}). 
The data were recorded using the Mark III recording system in mode B, 
with an effective bandwidth of 28 MHz centred at 4973 MHz.
Left circular polarization was recorded. 
Since the motion of some of the antennas is limited in hour angle,
we inevitably had to schedule a few scans 
when the source could not be observed at one or two telescopes. 
All sources were observed for three 
scans of 13 minutes (13$^m$ corresponds to a single pass on a tape).

The data were correlated using the VLBA correlator in Socorro, New Mexico,
four months after the observations took place. 
The output of the correlator provides a measure of the complex
fringe visibility sampled at intervals of 2 seconds on each baseline, 
at $7\times16$ frequencies within the 28 MHz band, with the phase
referenced to an {\it a priori} model of the source position, antenna 
locations, and atmosphere. The residual phase gradients in time and frequency
due to delay and rate errors in the {\it a priori} model are estimated and removed, during the process of ``fringe fitting''. Fringe fitting was performed 
using the AIPS task FRING, an implementation of the Schwab \& Cotton (1983) 
algorithm.
A solution interval of 3 minutes and a point source model were used, and
Effelsberg was taken as the ``reference telescope'' whenever possible.
No fringes were found for the Cambridge telescope.
The amplitude calibration was performed with the AIPS tasks ANTAB and
APCAL, using system temperature and antenna gain information. 
The visibility data were averaged across the observing band and then 
written in one single $u,v$-file per object.
The typical $u,v$ coverage obtained for a source is shown in figure \ref{UV}.

The final images were produced after several cycles of imaging and 
self-calibration 
using the AIPS tasks IMAGR and CALIB. Solution intervals were decreased in
each step, starting with a few minutes, until no increase of the image 
quality 
(using noise-level and the presence of negative structure as criteria) was 
detected. If a source was sufficiently strong, 
antenna amplitude solutions were also determined for each scan.
For each source a ``natural'' weighted image was produced.
If the $u,v$-data were of sufficient quality, a ``uniform'' weighted image 
was also produced.

\subsection{The 15 GHz Observations, Correlation and Reduction}

\begin{figure*}
\psfig{figure=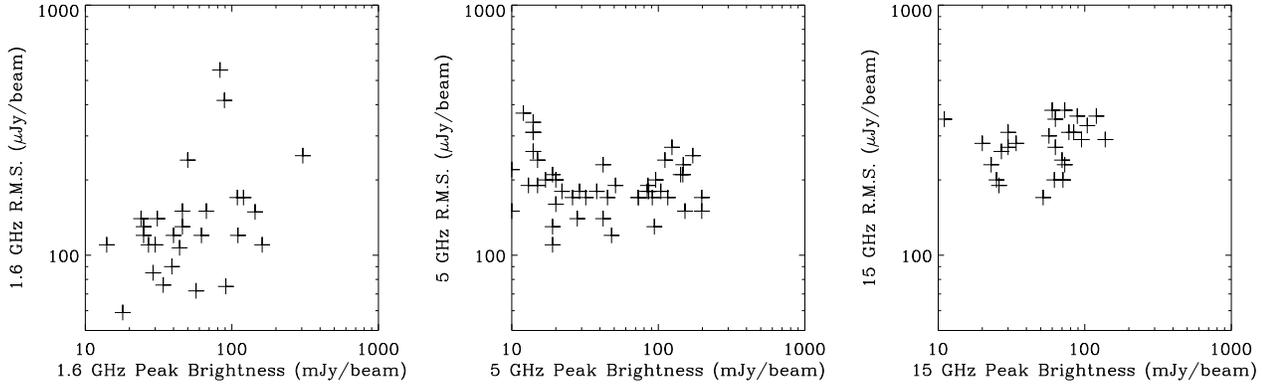,width=17cm}
\caption{\label{rmspeak} The rms noise levels as function of peak brightness
for the 1.6, 5 and 15 GHz images.}
\end{figure*}

The 15 GHz data were obtained during a 24 hour observing session on 
29 June 1996, using the ten telescopes of the VLBA. 
The data were recorded in $128-8-1$  mode (128 Mbits/sec, 8 IF channels,
1 bit/sample), with an effective bandwidth of 32 MHz centred at 15360 MHz.
All 27 sources in the sample with peak frequencies higher than 5 GHz and/or 
peak flux densities greater than 125 mJy were observed.
The expected maximum brightness in each of the images at 15 GHz was estimated 
from the overall radio spectra of the sources and their 5 GHz VLBI morphology.
In order to use the conventional fringe-fitting methods of VLBI imaging,
 the signal to noise ratio on each baseline within the coherence time has to 
be sufficiently high.
Sources with an expected maximum brightness at 15 GHz of $>60$ mJy/beam 
are sufficiently strong and were 
observed for 3 scans of 11 minutes each. However, sources with expected 
maximum brightnesses of $<60$ mJy/beam, were observed using a
``phase-referencing'' method to increase the signal to noise ratio.
This involves observations of the target source 
interspersed with observations of a nearby ($<2.5^{\circ}$) 
compact calibrator source. Measurements of residual delay and rate are made 
towards this bright source and transferred to the target source data.
We used cycles of 3 minutes on the target source and 1.5 minutes on the 
calibrator source. The total integration time on a target was 45 minutes 
divided over three scans. The sources for which the phase referencing technique
was required and the calibration sources used (in brackets) were 
B0400+6042 (B0354+599), B0436+6152 (B0444+634), B0513+7129 (B0518+705),
B0531+6121 (B0539+6200), B0538+7131 (B0535+6743), B0755+6354 (B0752+6355),\\
B1525+6801 (B1526+670), B1538+5920 (B1550+5815), B1600+7131 (B1531+722), 
B1819+6707 (B1842+681),\\ and B1841+6715 (B1842+681). 
Data reduction of the phase referenced observations is similar to
that for the 5 GHz data. The typical $u,v$ coverage obtained for a source at 
15 GHz is shown in figure \ref{UV}.

\subsection{The 1.6 GHz Observations, Correlation and Reduction}

The 1.6 GHz data were obtained during two 
observing sessions, both involving the ten telescopes of the VLBA and 4 
antennas of the EVN (see table \ref{tel}). The Westerbork data in the 
second session was lost due to technical failure.
The data were recorded in $128-4-2$  mode (128 Mbits/sec, 4 IF channels,
2 bit/sample), with an effective bandwidth of 32 MHz centred at 1663 MHz and
1655 during the first and second session respectively.
In the first session, a subsample of 23 objects was observed for
$2\times12$ hours on 14 and 16 September 1997. This subsample contained
all sources with peak frequencies $<5$ GHz, which were found to be extended in
the 5 GHz observations. In the second session, all 9 remaining sources 
with peak frequencies $<3$ GHz, which had not been imaged before at this 
frequency, were observed. The sources were typically observed for $4\times11$ 
minutes each, and an example of a $u,v$ coverage is shown in figure \ref{UV}.
The data were correlated in Socorro. No fringes were found for
B0513+7129, B0537+6444, and B0544+5847. 
Several sources in the second session were observed using phase referencing.
These sources, with their calibrators in brackets, are 
B0537+6444 (B0535+677), B0830+5813 (B0806+573), B1557+6220 (B1558+595), 
B1639+6711 (B1700+685), and B1808+6813 (B1749+701).
The data were reduced in a similar way as the data at 5 GHz.

\section{Results}

\begin{figure}
\psfig{figure=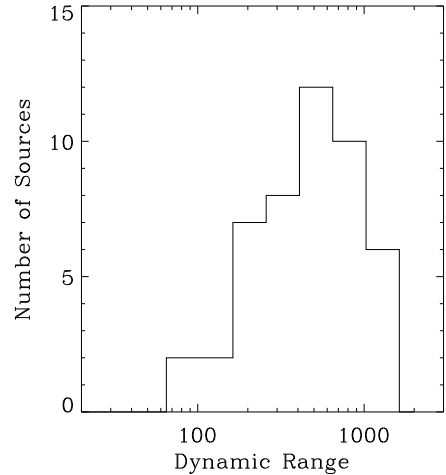,width=7cm}
\caption{\label{dynrange} The dynamic ranges for all sources in the sample at
the observed frequency closest to their spectral peak.}
\end{figure}

The parameters of the resulting 102 images (29 at 1.6 GHz, 47 at 5 GHz, and 
26 at 15 GHz) are given in table \ref{mappar}. 
Figure \ref{rmspeak} shows the rms noise as function of the peak brightness
in the images at the three observing frequencies. 
The dynamic ranges (defined as the ratio of the maximum brightness in the image
to the rms noise in an area of blank sky) are between 125 and 2500 at 1.6 GHz,
between 25 and 1700 at 5 GHz, and between 30 and 500 at 15 GHz.
At 1.6 GHz, two of the bright sources have higher rms-noise levels than
expected, which may indicate that the dynamic range is not limited by 
the thermal noise. To be able to compare the VLBI observations of this faint
sample with those on bright GPS samples, it is important to determine
whether components have been missed due to the limited dynamic range
for this faint sample. We therefore plotted the distribution of dynamic range
for the observations closest in frequency to the spectral peak 
(Fig. \ref{dynrange}).
Only 2 objects (B0755+6354 and B0544+5847) turn out not to have an image 
with a dynamic range $>100$.

\begin{figure*}
\psfig{figure=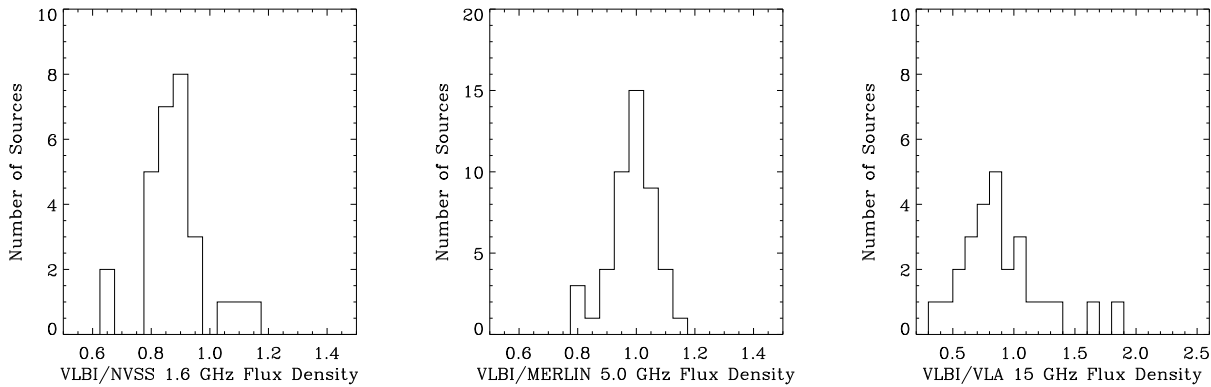,width=17cm}
\caption{\label{totvlbi} The ratio of total VLBI flux density in the maps
to the flux density in the NVSS at 1.6 GHz (left), 
to the MERLIN observations at 
5 GHz (middle), and to the VLA 15 GHz flux densities (right). }
\end{figure*}

In Figure \ref{totvlbi} the ratio of total VLBI flux density in the images
to the flux density in the NVSS at 1.6 GHz, to the MERLIN observations at 
5 GHz, and to the VLA 15 GHz flux densities (from Snellen et al. 1998a), are 
plotted.
This enables us to judge whether substantial structure has been
resolved out in the VLBI observations. At 1.6 GHz, typically 90\% of the NVSS 
flux density is recovered in the VLBI observations, while at 5 GHz the 
distribution peaks at 100\%. Only at 15 GHz is the distribution much broader
and peaks at about 80\% of the flux density in the VLA observations, and
hence provides some evidence that at this frequency some extended structure
may be missed. The broadness of the peak is probably also influenced
by variability.

Figures \ref{fig3}, \ref{fig2}, \ref{fig1} give the maps of 
the individual sources with observations at three, two and one 
frequency respectively. For each source, the images have the same size at 
each frequency, and are centred in such a way that identical components at 
different frequencies match in relative position.

\subsection{Model Fitting \label{modelfit}}

Quantitative parameters of the source brightness distributions were 
estimated by fitting elliptical Gaussian functions to the maps, using the 
AIPS-task JMFIT. 
For some of the complex sources it was necessary to restrict the
fit to a number of point sources (e.g. 0513+7129 at 5 GHz). 
In a few cases, the positions of some of the fitted components were
kept fixed to correspond to their positions at higher frequency 
(e.g. 0752+6355 at 5 GHz). We checked whether 
the model was a good representation of the source structure, by
comparing the total flux density in the image to that in the model, and 
by ensuring that the residual image did not 
show any significant negative structure.
A spectral decomposition was performed by matching the components believed
to correspond to each other at the different frequencies. 
Due to the increase in resolution with frequency, some components at the
higher frequencies were combined to match a single component at 
the lower frequency. The decomposed spectra are shown along with the images 
in 
figures \ref{fig3} and \ref{fig2}.
The results of the fits are given in table \ref{sourcepar}.

Column 1 gives the source name, column 2 the figure in which the maps are 
shown, column 3 the classification (as discussed in the next section), 
and in column 4 the component name used for the spectral decomposition.
Columns 5 to 9 give for each component observed at 1.6 GHz the flux 
density, relative position in R.A. and Dec., and the fitted angular size 
(major and minor axis, and position angle). Columns 10 to 14, and 
columns 15 to 19 give the same for the components observed at
5 GHz and 15 GHz respectively.

\subsection{Classification of the Radio Morphologies}

We classify the radio morphologies in four ways:
\begin{itemize}
\item[1)] {\bf Compact Symmetric Objects (CSO)}. Sources with a compact flat 
spectrum component with extended components with steeper 
spectra on either side.
\item[2)] {\bf Core-Jet sources (CJ)}. Sources with compact flat spectrum 
component with one or more components with steeper spectra on one side only.
\item[3)] {\bf Compact Double (CD)}. Sources showing two dominant 
components with comparable spectra, but no evidence of a central flat spectrum
component.
\item[4)] {\bf Complex sources (CX)}. Sources with a complex morphology, not
falling in one of the above categories.
\end{itemize}

From the 47 sources in the sample, 3 could be classified as CSO,
11 as CD, 7 as CJ, and 2 as CX. Of the 25 remaining sources, 2 were  
resolved at only one frequency, 2 were only observed at one frequency, 
and 20 show a single component at 2 frequencies, and therefore could
not be classified. For one source
(1642+6701) it was not clear how to overlay the 1.6 and 5 GHz maps.
The individual sources are briefly discussed below.

\subsubsection{Discussion of Individual Sources}

\noindent {\bf B0400+6042: CD}
Several components are visible with the two outer components having 
comparable spectra and the central component 
having a marginally flatter spectrum. This source is tentatively classified as 
a CD, but it could also be a CSO.

\noindent {\bf B0436+6152: CSO}
The brightest component at 15 GHz in the centre is interpreted as 
the core, with extended steeper spectrum components to the north-east
and south-west. This source is classified as a CSO. 

\noindent {\bf B0513+7129: CX}
This object shows two dominant components with a jet-like feature pointed
to the north-west. Although the component with the flattest 
spectrum is located in the center, this source is classified as 
a complex source due to the strangely bent structure.

\noindent {\bf B0539+6200: CJ}
The south-western component has a flatter spectral index than the 
north-eastern component. Only two components are visible, hence this object 
is tentatively classified as a CJ. 

\noindent {\bf B0752+6355: CX}
The ``C'' shaped morphology of this compact source showing components 
with a large range of spectral indices, leads us
to classify this object as one with a complex morphology.

\noindent {\bf B1620+6406: CD}
The steep spectrum of the northern component is similar to the 
spectrum of the southern component (using the upper-limit for the flux
density at 5 GHz). We therefore tentatively classify this object as a CD.

\noindent {\bf B1642+6701: -}
It was not possible to reliably overlay the two maps at 1.6 and 5 GHz, 
but the most likely match is shown in figure \ref{fig2}.
Due to this uncertainty, it was not possible to reliably classify this 
source.

\noindent {\bf B1647+6225: CD}
The two components have similar spectra, with a possible jet leading
to the northern component. This object is tentatively classified as a CD.

\noindent {\bf B1655+6446: CD}
Only the southern component is detected at 5 GHz. However its steep spectral 
index, and the upper-limit to the spectral index of the northern component
makes us tentatively classify this source as a CD. 

\noindent {\bf B1657+5826: CD}
Only the western component is detected at 5 GHz. However its steep spectral 
index, and the upper-limit to the spectral index of the eastern component
makes us classify this source as a CD. 

\noindent {\bf B1819+6707: CSO}
Two dominant components are visible in this source at 1.6, 5, and 15 GHz 
with comparable spectra. A faint compact component is visible in between
in the 5 GHz map. This object is therefore classified as a CSO.

\noindent {\bf B1942+7214: CJ}
This object shows faint extended structure to the south-west in its 
1.6 and 5 GHz images. The bright northern component appears to
have a flatter spectrum than the faint extended emission. 
We tentatively classify this object as a core-jet.

\noindent {\bf B1946+7048: CSO}
This source is the archetype compact symmetric object (CSO), and has
been discussed in detail by Taylor and Vermeulen (1997).
The core is only visible at 15 GHz.

\noindent {\bf B1954+6146: CJ}
Only the flat spectrum compact component in the south is detected at 
15 GHz. The limit to the spectral index of the northern component makes
us tentatively classify this source as a core-jet.

\begin{table}
\begin{tabular}{crrrrrrrrrrrr}
Name&\multicolumn{4}{c}{1.6 GHz Map Parameters}&\multicolumn{4}{c}{5 GHz Map Parameters}&\multicolumn{4}{c}{15 GHz Map Parameters}\\
&\multicolumn{2}{c}{Beam}&rms&$S_{peak}$&\multicolumn{2}{c}{Beam}&rms&$S_{peak}$&\multicolumn{2}{c}{Beam}&rms&$S_{peak}$\\
&(mas)&$^{\circ}$&$\frac{\mu Jy}{beam}$&$\frac{mJy}{beam}$&(mas)&$^{\circ}$&$\frac{\mu Jy}{beam}$&$\frac{mJy}{beam}$&(mas)&$^{\circ}$&$\frac{\mu Jy}{beam}$&$\frac{mJy}{beam}$\\
0400+6042&&&&&$1.2\times1.1$&65.8&180&38&$1.1\times0.6$&-22.4&280&34\\
0436+6152&$5.6\times2.5$&-26.3&120&110&$3.4\times1.2$&68.5&230&42&$1.3\times1.1$&-35.3&200&25\\
0441+5757&&&&&$2.3\times1.3$&70.5&200&96&$1.6\times1.2$&-70.2&230&73\\
0513+7129&&&&&$1.4\times0.9$&14.2&190&51&$0.8\times0.7$&41.0&170&52\\
0531+6121&&&&&$2.0\times1.0$&24.0&130&19&$1.2\times1.0$&-46.4&270&30\\
0535+6743&$9.7\times 3.7$&-5.8&416&89&$1.6\times1.0$&10.3&230&148&$1.4\times1.2$&45.9&200&71\\
0537+6444&$9.3\times 3.7$&-10.6&85&29&$2.2\times1.0$&12.8 &190&13&&&\\
0538+7131&&&&&$1.5\times1.1$&13.4&170&73&$1.9\times1.6$&-51.5&310&30\\
0539+6200&$4.4\times2.9$&-2.6&550&83&$1.9\times1.0$&22.4&170&91&$0.8\times0.5$&56.9&290&95\\
0543+6523&$3.8\times3.0$&-11.6&240&50&$1.7\times1.1$&23.1 &170&32&&&&\\
0544+5847&&&&&$1.9\times1.0$&-7.1 &210&19&&&\\
0552+6017&$4.3\times2.9$&-7.4&120&40&$2.1\times0.9$&3.4 &350&8&&&&\\
0557+5717&$4.7\times3.0$&3.7&90&39&$1.4\times0.9$&-11.7 &150&10&&&&\\
0601+5753&&&&&$1.5\times1.0$&-1.5&210&147&$1.5\times0.9$&-44.7&380&73\\
0748+6343&&&&&$1.5\times1.0$&15.1 &270&124&&&\\
0752+6355&&&&&$1.0\times0.9$&19.1&150&152&$0.7\times0.5$&-18.4&330&104\\
0755+6354&&&&&$2.8\times1.9$&12.8&340&14&$0.7\times0.5$&11.8&280&20\\
0756+6647&&&&&$2.2\times1.1$&46.3&130&94&$0.7\times0.5$&-33.5&270&63\\
0758+5929&$4.6\times 2.6$&-18.2&149&144&$2.5\times1.2$&20.1&170&116&$1.3\times0.5$&13.5&200&62\\
0759+6557&$4.1\times2.9$&-4.7&140&24&$1.9\times1.1$&-27.1 &220&10&&&&\\
0826+7045&&&&&$1.3\times1.2$&15.1 &190&85&&&\\
0830+5813&$4.3\times 2.5$&-15.7&107&44&$1.6\times1.1$&9.0 &180&29&&&\\
1525+6801&$4.3\times3.2$&-72.0&170&109&$1.1\times1.0$&7.0&120&48&$1.4\times1.4$&-29.8&230&23\\
1538+5920&&&&&$2.7\times1.2$&0.4&140&42&$0.5\times0.5$&0.0&350&11\\
1550+5815&&&&&$2.6\times1.2$&-0.4&170&198&$1.2\times0.4$&15.8&360&120\\
1551+6822&$4.4\times3.2$&-65.4&130&46&$3.3\times1.3$&6.2 &160&20&&&&\\
1557+6220&$4.4\times 2.8$&6.6&76&34&$2.7\times1.1$&-2.8 &240&15&&&\\
1600+7131&$3.9\times3.1$&64.3&110&161&$2.3\times1.5$&68.6&170&45&$0.6\times0.5$&-40.0&190&26\\
1620+6406&$3.9\times3.3$&63.3&110&27&$3.6\times3.3$&-33.4 &370&12&&&&\\
1622+6630&&&&&$2.9\times1.2$&-28.7&150&198&$0.8\times0.5$&16.6&290&138\\
1639+6711&$4.1\times 2.7$&13.3&72&57&$2.1\times1.1$&0.1 &140&28&&&\\
1642+6701&$3.9\times3.2$&72.2&120&62&$1.4\times0.9$&13.9 &110&19&&&&\\
1647+6225&$4.0\times3.3$&36.5&130&25&$3.6\times3.1$&-33.2 &190&15&&&&\\
1655+6446&$3.7\times3.2$&1.6&140&31&$3.1\times2.0$&-28.5 &310&14&&&&\\
1657+5826&$3.9\times3.2$&1.0&110&30&$3.8\times2.1$&38.0 &260&14&&&&\\
1746+6921&$5.3\times 3.0$&20.6&75&91&$1.5\times1.0$&2.1&180&86&$1.0\times0.7$&-76.8&350&63\\
1807+5959&$4.1\times3.4$&-89.8&110&14&$2.5\times1.0$&3.2 &170&26&&&&\\
1807+6742&$3.9\times3.2$&80.4&120&25&$4.8\times3.6$&50.1 &310&14&&&&\\
1808+6813&$4.1\times 2.8$&19.9&59&18&$2.6\times0.8$&-4.1 &200&17&&&\\
1819+6707&$3.9\times3.2$&57.3&150&67&$2.5\times1.1$&46.1&180&22&$0.9\times0.7$&-21.7&260&27\\
1841+6715&$3.9\times3.2$&84.4&170&120&$2.8\times1.1$&35.0&180&104&$1.3\times1.0$&15.1&300&57\\
1843+6305&$4.0\times3.3$&83.3&150&46&$3.0\times0.9$&9.7 &200&20&&&&\\
1942+7214&$3.9\times 2.8$&33.5&65&161&$3.3\times1.6$&15.2&250&172&$0.7\times0.5$&23.2&310&84\\
1945+6024&&&&&$2.7\times1.1$&-3.7&170&73&$0.9\times0.5$&22.8&310&78\\
1946+7048&$3.9\times3.3$&-79.7&250&305&$1.0\times0.9$&-28.6&180&84&$1.6\times1.1$&28.6&240&70\\
1954+6146&&&&&$2.5\times1.2$&-1.6&210&142&$0.8\times0.5$&2.9&360&89\\
1958+6158&&&&&$2.5\times1.1$&0.0&240&111&$1.2\times0.6$&-13.4&380&60\\ 
\end{tabular}
\caption{ \label{mappar} Relevant parameters of the presented maps.}
\end{table}

\begin{table*}
\setlength{\tabcolsep}{1mm}
\begin{tabular}{ccrc|rrrcr|rrrcr|rrrcr}
Name&Fg&Cls&Cp&\multicolumn{5}{c}{1.6 GHz Data}&\multicolumn{5}{c}{5.0 GHz Data}&\multicolumn{5}{c}{14.9 GHz Data}\\
&&&&Flx&$\Delta$X &$\Delta$Y&Size&PA&Flx&$\Delta$X&$\Delta$Y&Size&PA&Flx&$\Delta$X&$\Delta$Y&Size&PA\\
&&&&mJy&mas&mas&mas&$^\circ$&mJy&mas&mas&mas&$^\circ$&mJy&mas&mas&mas&$^\circ$\\
0400+6042&2&CD&
E&
&&&&&
$ 62.7$$\pm$$3.1$&  0.0&  0.0&$1.8$$\times$$1.2$&$140$&
$ 58.5$$\pm$$2.9$&  0.0&  0.0&$1.3$$\times$$0.2$&$153$
\\
&&&
W&
&&&&&
$ 11.0$$\pm$$0.6$&  4.1&  1.7&$2.2$$\times$$1.7$&$ 71$&
$  1.9$$\pm$$0.2$&  4.1&  1.7&$               -$&$  0$
\\
&&&
W&
&&&&&
$  2.8$$\pm$$0.2$&  1.2&  1.2&$            <0.2$&$  0$&
$  6.1$$\pm$$0.4$&  1.2&  1.2&$               -$&$  0$
\\
&&&
W&
&&&&&
$  3.3$$\pm$$0.3$&  2.0&  1.2&$            <0.1$&$  0$&
$  2.3$$\pm$$0.2$&  2.0&  1.2&$               -$&$  0$
\\
0436+6152&1&CSO&
S&
$169$$\pm$$ 8$&  0.0&  0.0&$3.1$$\times$$0.9$&$ 41$&
$ 84.7$$\pm$$4.2$&  0.0&  0.0&$2.9$$\times$$0.9$&$ 38$&
$  7.2$$\pm$$0.4$&  0.0&  0.0&$2.7$$\times$$0.8$&$ 41$
\\
&&&
C&
$ 36.8$$\pm$$1.9$& -2.8&  4.8&$1.5$$\times$$0.6$&$ 67$&
$ 37.3$$\pm$$1.9$& -3.2&  5.0&$1.4$$\times$$0.8$&$ 64$&
$ 25.5$$\pm$$1.3$& -3.0&  4.8&$0.2$$\times$$0.1$&$ 50$
\\
&&&
C&
&&&&&
&&&&&
$  2.2$$\pm$$0.2$& -3.7&  6.4&$1.5$$\times$$0.5$&$ 43$
\\
&&&
C&
&&&&&
&&&&&
$  2.8$$\pm$$0.2$& -1.8&  3.2&$1.2$$\times$$0.8$&$ 66$
\\
&&&
N&
$  3.1$$\pm$$0.3$& -5.2& 10.5&$            <1.2$&$  0$&
$  2.2$$\pm$$0.2$& -5.9& 11.4&$1.8$$\times$$0.8$&$ 50$&
$  4.4$$\pm$$0.3$& -5.7&  9.5&$2.0$$\times$$0.5$&$ 32$
\\
&&&
N&
$  4.3$$\pm$$0.3$& -8.4& 14.9&$9.3$$\times$$2.9$&$ 21$&
&&&&&
&&&&
\\
0441+5757&2&$-$&
C&
&&&&&
$111$$\pm$$ 5$&  0.0&  0.0&$1.0$$\times$$0.5$&$ 45$&
$ 73.9$$\pm$$3.7$&  0.0&  0.0&$            <0.1$&$  0$
\\
0513+7129&2&CX&
N&
&&&&&
$ 64.7$$\pm$$3.2$&  0.0&  0.0&$0.8$$\times$$0.5$&$ 34$&
$ 54.7$$\pm$$2.7$&  0.0&  0.0&$            <0.1$&$  0$
\\
&&&
S&
&&&&&
$ 60.4$$\pm$$3.0$& -0.7& -3.5&$1.2$$\times$$0.6$&$ 33$&
$ 18.6$$\pm$$1.0$& -0.6& -3.6&$            <0.1$&$  0$
\\
&&&
W&
&&&&&
$  4.2$$\pm$$0.3$&  2.0&  1.1&$               -$&$  0$&
&&&&
\\
&&&
W&
&&&&&
$  2.0$$\pm$$0.2$&  2.9&  1.4&$               -$&$  0$&
&&&&
\\
&&&
W&
&&&&&
$  3.7$$\pm$$0.3$&  4.8&  2.3&$               -$&$  0$&
$  2.5$$\pm$$0.2$&  5.1&  1.8&$1.5$$\times$$0.6$&$ 65$
\\
&&&
W&
&&&&&
$  2.6$$\pm$$0.2$&  6.4&  3.4&$               -$&$  0$&
&&&&
\\
0531+6121&2&$-$&
C&
&&&&&
$ 18.7$$\pm$$1.0$&  0.0&  0.0&$            <0.2$&$  0$&
$ 30.1$$\pm$$1.5$&  0.0&  0.0&$0.2$$\times$$0.1$&$142$
\\
0535+6743&1&$-$&
C&
$ 94.5$$\pm$$4.7$&  0.0&  0.0&$4.0$$\times$$1.9$&$165$&
$166$$\pm$$ 8$&  0.0&  0.0&$0.5$$\times$$0.4$&$174$&
$ 78.8$$\pm$$3.9$&  0.0&  0.0&$0.5$$\times$$0.4$&$ 94$
\\
0537+6444&2&$-$&
C&
$ 31.9$$\pm$$1.6$&  0.0&  0.0&$1.8$$\times$$0.9$&$  9$&
$ 15.3$$\pm$$0.8$&  0.0&  0.0&$1.3$$\times$$0.2$&$ 12$&
&&&&
\\
&&&
C&
&&&&&
$  1.3$$\pm$$0.2$&  1.1& -4.7&$               -$&$  0$&
&&&&
\\
0538+7131&2&$-$&
C&
&&&&&
$ 78.0$$\pm$$3.9$&  0.0&  0.0&$0.5$$\times$$0.3$&$152$&
$ 29.6$$\pm$$1.5$&  0.0&  0.0&$            <0.2$&$  0$
\\
0539+6200&1&CJ&
W&
$ 96.8$$\pm$$4.8$&  0.0&  0.0&$1.7$$\times$$1.2$&$126$&
$101$$\pm$$ 5$&  0.0&  0.0&$0.6$$\times$$0.4$&$163$&
$ 95.0$$\pm$$4.8$&  0.0&  0.0&$            <0.2$&$  0$
\\
&&&
E&
$ 10.3$$\pm$$0.6$& -4.6&  4.0&$2.7$$\times$$0.8$&$121$&
$  4.3$$\pm$$0.3$& -4.8&  2.8&$3.3$$\times$$1.8$&$ 36$&
&&&&
\\
0543+6523&2&$-$&
C&
$ 62.7$$\pm$$3.1$&  0.0&  0.0&$2.1$$\times$$1.3$&$ 90$&
$ 36.8$$\pm$$1.9$&  0.0&  0.0&$0.7$$\times$$0.5$&$ 36$&
&&&&
\\
&&&
C&
&&&&&
$  5.2$$\pm$$0.3$&  2.5& -0.6&$1.2$$\times$$0.5$&$ 36$&
&&&&
\\
0544+5847&3&$-$&
C&
&&&&&
$ 18.6$$\pm$$1.0$&  0.0&  0.0&$0.4$$\times$$0.1$&$175$&
&&&&
\\
&&&
C&
&&&&&
$  7.9$$\pm$$0.4$&  0.6&  1.6&$0.7$$\times$$0.4$&$ 21$&
&&&&
\\
&&&
C&
&&&&&
$  1.4$$\pm$$0.2$&  1.3&  4.7&$            <1.1$&$  0$&
&&&&
\\
&&&
C&
&&&&&
$  3.0$$\pm$$0.2$&  1.6&  8.0&$1.8$$\times$$0.7$&$173$&
&&&&
\\
0552+6017&2&$-$&
C&
$ 41.8$$\pm$$2.1$&  0.0&  0.0&$0.9$$\times$$0.2$&$ 71$&
$ 13.6$$\pm$$0.7$&  0.0&  0.0&$1.5$$\times$$0.3$&$114$&
&&&&
\\
&&&
C&
$  1.2$$\pm$$0.2$&-12.2& -3.3&$            <1.7$&$  0$&
&&&&&
&&&&
\\
0557+5717&2&CJ&
N&
$ 49.2$$\pm$$2.5$&  0.0&  0.0&$2.5$$\times$$1.2$&$ 44$&
$ 12.5$$\pm$$0.7$&  0.0&  0.0&$3.1$$\times$$1.5$&$ 42$&
&&&&
\\
&&&
S&
$ 10.9$$\pm$$0.6$&  3.0& -5.8&$2.3$$\times$$1.5$&$124$&
$ 16.3$$\pm$$0.8$&  2.2& -8.9&$3.4$$\times$$1.8$&$ 18$&
&&&&
\\
0601+5753&2&$-$&
C&
&&&&&
$188$$\pm$$ 9$&  0.0&  0.0&$0.8$$\times$$0.3$&$106$&
$ 94.9$$\pm$$4.7$&  0.0&  0.0&$0.9$$\times$$0.3$&$ 98$
\\
0748+6343&3&$-$&
C&
&&&&&
$126$$\pm$$ 6$&  0.0&  0.0&$0.2$$\times$$0.1$&$ 65$&
&&&&
\\
0752+6355&2&CX&
C&
&&&&&
$182$$\pm$$ 9$&  0.0&  0.0&$1.2$$\times$$0.2$&$167$&
$ 31.6$$\pm$$1.6$&  0.0&  0.0&$0.4$$\times$$0.1$&$144$
\\
&&&
S&
&&&&&
$105$$\pm$$ 5$&  0.3& -0.5&$1.9$$\times$$0.5$&$ 24$&
$ 62.3$$\pm$$3.1$&  0.3& -0.5&$0.8$$\times$$0.3$&$ 24$
\\
&&&
N&
&&&&&
$ 17.4$$\pm$$0.9$&  0.3&  0.7&$            <0.8$&$  0$&
$108$$\pm$$ 5$&  0.3&  0.7&$0.2$$\times$$0.1$&$142$
\\
&&&
W&
&&&&&
$  2.4$$\pm$$0.2$&  1.8&  1.5&$               -$&$  0$&
&&&&
\\
&&&
W&
&&&&&
$  4.6$$\pm$$0.3$&  4.2&  1.8&$               -$&$  0$&
&&&&
\\
0755+6354&2&$-$&
C&
&&&&&
$ 19.9$$\pm$$1.0$&  0.0&  0.0&$0.7$$\times$$0.2$&$132$&
$ 19.9$$\pm$$1.0$&  0.0&  0.0&$            <0.1$&$  0$
\\
0756+6647&2&$-$&
C&
&&&&&
$ 98.1$$\pm$$4.9$&  0.0&  0.0&$0.4$$\times$$0.3$&$ 34$&
$ 72.6$$\pm$$3.6$&  0.0&  0.0&$0.3$$\times$$0.2$&$ 29$
\\
0758+5929&1&CD&
W&
$122$$\pm$$ 6$&  0.0&  0.0&$               -$&$  0$&
$121$$\pm$$ 6$&  0.0&  0.0&$0.8$$\times$$0.2$&$ 36$&
$ 76.4$$\pm$$3.8$&  0.0&  0.0&$0.7$$\times$$0.2$&$ 31$
\\
&&&
E&
$ 71.4$$\pm$$3.6$& -1.7&  1.0&$               -$&$  0$&
$ 34.1$$\pm$$1.7$& -1.8&  0.9&$0.6$$\times$$0.5$&$ 41$&
$ 10.4$$\pm$$0.6$& -1.9&  0.8&$0.6$$\times$$0.3$&$ 12$
\\
0759+6557&2&CD&
W&
$ 25.5$$\pm$$1.3$&  0.0&  0.0&$0.9$$\times$$0.6$&$  3$&
$ 12.6$$\pm$$0.7$&  0.0&  0.0&$0.9$$\times$$0.6$&$154$&
&&&&
\\
&&&
E&
$ 17.8$$\pm$$0.9$& -4.8& -5.6&$1.7$$\times$$1.4$&$ 62$&
$  4.1$$\pm$$0.3$& -5.5& -5.8&$1.3$$\times$$1.0$&$ 23$&
&&&&
\\
&&&
E&
&&&&&
$  1.8$$\pm$$0.2$& -3.0& -5.3&$1.7$$\times$$1.4$&$142$&
&&&&
\\
0826+7045&3&$-$&
C&
&&&&&
$ 90.5$$\pm$$4.5$&  0.0&  0.0&$0.4$$\times$$0.2$&$162$&
&&&&
\\
0830+5813&2&$-$&
C&
$ 50.8$$\pm$$2.5$&  0.0&  0.0&$1.9$$\times$$1.0$&$  0$&
$ 39.6$$\pm$$2.0$&  0.0&  0.0&$0.6$$\times$$0.5$&$ 66$&
&&&&
\\
1525+6801&1&CD&
S&
$137$$\pm$$ 6$&  0.0&  0.0&$2.5$$\times$$1.1$&$170$&
$ 80.9$$\pm$$4.0$&  0.0&  0.0&$1.1$$\times$$0.6$&$151$&
$ 24.4$$\pm$$1.2$&  0.0&  0.0&$0.4$$\times$$0.2$&$171$
\\
&&&
N&
$ 14.5$$\pm$$0.8$& 13.1& 17.8&$2.1$$\times$$1.2$&$166$&
$  8.5$$\pm$$0.5$& 13.2& 18.1&$1.5$$\times$$0.8$&$170$&
$  3.9$$\pm$$0.3$& 12.7& 18.4&$2.9$$\times$$0.9$&$ 22$
\\
&&&
N&
$ 15.6$$\pm$$0.8$& 10.2& 13.7&$5.4$$\times$$3.4$&$135$&
$  5.8$$\pm$$0.4$& 11.5& 16.3&$2.4$$\times$$1.9$&$112$&
&&&&
\\
&&&
N&
&&&&&
$  2.3$$\pm$$0.2$& 10.4& 13.4&$1.6$$\times$$0.3$&$ 36$&
&&&&
\\
1538+5920&2&$-$&
C&
&&&&&
$ 45.9$$\pm$$2.3$&  0.0&  0.0&$0.7$$\times$$0.3$&$ 27$&
$ 20.1$$\pm$$1.0$&  0.0&  0.0&$0.7$$\times$$0.3$&$ 20$
\\
1550+5815&2&$-$&
C&
&&&&&
$233$$\pm$$11$&  0.0&  0.0&$1.4$$\times$$0.2$&$158$&
$174$$\pm$$ 8$&  0.0&  0.0&$0.7$$\times$$0.2$&$164$
\\
1551+6822&2&$-$&
C&
$ 49.6$$\pm$$2.5$&  0.0&  0.0&$1.7$$\times$$0.5$&$105$&
$ 21.1$$\pm$$1.1$&  0.0&  0.0&$            <0.3$&$  0$&
&&&&
\\
&&&
C&
&&&&&
$  5.1$$\pm$$0.3$& -2.3& -0.9&$            <1.0$&$  0$&
&&&&
\\
1557+6220&2&$-$&
C&
$ 39.3$$\pm$$2.0$&  0.0&  0.0&$1.5$$\times$$0.8$&$ 89$&
$ 17.9$$\pm$$0.9$&  0.0&  0.0&$1.2$$\times$$0.4$&$156$&
&&&&
\\
\end{tabular}
\caption{\label{sourcepar} The fitted parameters of the observed components.}
\end{table*}
\addtocounter{table}{-1}
\begin{table*}
\setlength{\tabcolsep}{1mm}
\begin{tabular}{ccrc|rrrcr|rrrcr|rrrcr}
Name&Fg&Cls&Cp&\multicolumn{5}{c}{1.6 GHz Data}&\multicolumn{5}{c}{5.0 GHz Data}&\multicolumn{5}{c}{14.9 GHz Data}\\
&&&&Flx&$\Delta$X&$\Delta$Y&Size&PA&Flx&$\Delta$X&$\Delta$Y&Size&PA&Flx&$\Delta$X&$\Delta$Y&Size&PA\\
&&&&mJy&mas&mas&mas&$^\circ$&mJy&mas&mas&mas&$^\circ$&mJy&mas&mas&mas&$^\circ$\\
1557+6220&2&$-$&
C&
$ 39.3$$\pm$$2.0$&  0.0&  0.0&$1.5$$\times$$0.8$&$ 89$&
$ 17.9$$\pm$$0.9$&  0.0&  0.0&$1.2$$\times$$0.4$&$156$&
&&&&
\\
1600+7131&1&CD&
N&
$261$$\pm$$13$&  0.0&  0.0&$3.1$$\times$$2.2$&$124$&
$ 54.4$$\pm$$2.7$&  0.0&  0.0&$1.0$$\times$$0.7$&$ 47$&
$ 26.3$$\pm$$1.3$&  0.0&  0.0&$            <0.1$&$  0$
\\
&&&
N&
&&&&&
$ 19.8$$\pm$$1.0$& -1.4& -1.7&$2.4$$\times$$1.4$&$110$&
&&&&
\\
&&&
N&
&&&&&
$  4.0$$\pm$$0.3$&  2.6& -3.6&$            <1.4$&$  0$&
&&&&
\\
&&&
S&
$  8.3$$\pm$$0.5$&  4.7&-22.2&$2.2$$\times$$1.2$&$115$&
$  6.7$$\pm$$0.4$&  7.7&-20.9&$1.5$$\times$$1.2$&$141$&
&&&&
\\
&&&
S&
$ 15.9$$\pm$$0.8$&  8.2&-19.3&$1.9$$\times$$1.4$&$165$&
&&&&&
&&&&
\\
1620+6406&2&CD&
N&
$ 29.0$$\pm$$1.5$&  0.0&  0.0&$1.1$$\times$$0.8$&$142$&
$ 12.9$$\pm$$0.7$&  0.0&  0.0&$1.0$$\times$$0.8$&$170$&
&&&&
\\
&&&
S&
$  5.2$$\pm$$0.3$&  0.3&-14.2&$            <1.2$&$  0$&
&&&&&
&&&&
\\
1622+6630&2&$-$&
C&
&&&&&
$218$$\pm$$10$&  0.0&  0.0&$0.6$$\times$$0.2$&$ 60$&
$169$$\pm$$ 8$&  0.0&  0.0&$0.3$$\times$$0.2$&$122$
\\
1639+6711&2&$-$&
C&
$ 68.5$$\pm$$3.4$&  0.0&  0.0&$1.8$$\times$$0.9$&$ 92$&
$ 38.6$$\pm$$1.9$&  0.0&  0.0&$1.0$$\times$$0.7$&$ 69$&
&&&&
\\
1642+6701&2&$-$&
W&
$ 83.4$$\pm$$4.2$&  0.0&  0.0&$3.5$$\times$$0.5$&$ 86$&
$ 14.8$$\pm$$0.8$&  0.0&  0.0&$1.3$$\times$$0.3$&$ 76$&
&&&&
\\
&&&
C&
$  8.7$$\pm$$0.5$& -4.9&  0.0&$            <0.8$&$  0$&
$ 36.8$$\pm$$1.9$& -2.7&  0.3&$1.5$$\times$$0.4$&$ 87$&
&&&&
\\
&&&
E&
$  7.8$$\pm$$0.4$& -9.6&  0.1&$5.1$$\times$$0.9$&$ 99$&
$  9.8$$\pm$$0.5$& -5.8&  0.5&$2.1$$\times$$0.6$&$ 79$&
&&&&
\\
1647+6225&2&CD&
N&
$ 24.2$$\pm$$1.2$&  0.0&  0.0&$            <0.4$&$  0$&
$ 15.8$$\pm$$0.8$&  0.0&  0.0&$0.7$$\times$$0.5$&$ 65$&
&&&&
\\
&&&
N&
$  7.0$$\pm$$0.4$&  4.1& -2.0&$2.9$$\times$$1.0$&$ 81$&
&&&&&
&&&&
\\
&&&
S&
$ 16.1$$\pm$$0.8$& 10.0&-12.0&$1.9$$\times$$0.7$&$ 76$&
$  2.7$$\pm$$0.2$& 10.5&-12.3&$0.8$$\times$$0.3$&$ 19$&
&&&&
\\
1655+6446&2&CD&
S&
$ 38.5$$\pm$$1.9$&  0.0&  0.0&$2.2$$\times$$1.1$&$ 76$&
$ 18.2$$\pm$$0.9$&  0.0&  0.0&$2.0$$\times$$1.0$&$108$&
&&&&
\\
&&&
N2&
$ 10.2$$\pm$$0.5$& -4.7&  5.1&$            <2.1$&$  0$&
&&&&&
&&&&
\\
&&&
N&
$  2.6$$\pm$$0.2$&-19.8& 15.0&$2.1$$\times$$1.3$&$  0$&
&&&&&
&&&&
\\
1657+5826&2&CD&
S&
$ 34.6$$\pm$$1.7$&  0.0&  0.0&$1.4$$\times$$1.4$&$ 90$&
$ 20.2$$\pm$$1.0$&  0.0&  0.0&$1.9$$\times$$1.6$&$ 31$&
&&&&
\\
&&&
N&
$  4.2$$\pm$$0.3$&-27.0&  6.8&$            <2.8$&$  0$&
&&&&&
&&&&
\\
1746+6921&1&CJ&
W&
$ 67.7$$\pm$$3.4$&  0.0&  0.0&$2.7$$\times$$0.9$&$116$&
$102$$\pm$$ 5$&  0.0&  0.0&$               -$&$  0$&
$ 70.8$$\pm$$3.5$&  0.0&  0.0&$0.5$$\times$$0.1$&$105$
\\
&&&
C&
$ 67.7$$\pm$$3.4$&-64.0&-65.0&$               -$&$  0$&
$ 50.3$$\pm$$2.5$& -1.1& -0.5&$               -$&$  0$&
$ 12.5$$\pm$$0.7$& -1.2& -0.4&$1.4$$\times$$0.2$&$129$
\\
&&&
E&
$  2.5$$\pm$$0.2$& -7.4& -4.3&$7.6$$\times$$2.1$&$149$&
&&&&&
&&&&
\\
1807+5959&2&CJ&
S&
$ 27.2$$\pm$$1.4$&  0.0&  0.0&$1.1$$\times$$0.4$&$173$&
$ 29.3$$\pm$$1.5$&  0.0&  0.0&$1.1$$\times$$0.1$&$  8$&
&&&&
\\
&&&
C&
$  6.0$$\pm$$0.4$&  0.1&  4.8&$2.1$$\times$$1.9$&$ 39$&
$  1.5$$\pm$$0.2$& -0.2&  6.3&$            <0.8$&$  0$&
&&&&
\\
&&&
N&
$ 12.5$$\pm$$0.7$& -0.6& 12.4&$4.7$$\times$$1.4$&$  6$&
$  4.0$$\pm$$0.3$& -0.6& 13.0&$3.8$$\times$$0.8$&$  9$&
&&&&
\\
1807+6742&2&CJ&
S&
$ 26.5$$\pm$$1.3$&  0.0&  0.0&$1.1$$\times$$0.7$&$104$&
$  5.0$$\pm$$0.3$&  0.0&  0.0&$2.9$$\times$$2.2$&$165$&
&&&&
\\
&&&
N&
$ 11.5$$\pm$$0.6$&  1.1&  5.9&$2.6$$\times$$0.4$&$178$&
$ 14.8$$\pm$$0.8$&  1.1&  6.8&$            <0.7$&$  0$&
&&&&
\\
1808+6813&2&$-$&
C&
$ 26.9$$\pm$$1.4$&  0.0&  0.0&$3.5$$\times$$1.2$&$167$&
$ 19.0$$\pm$$1.0$&  0.0&  0.0&$            <0.7$&$  0$&
&&&&
\\
1819+6707&1&CSO&
E&
$123$$\pm$$ 6$&  0.0&  0.0&$4.6$$\times$$2.2$&$ 58$&
$ 70.2$$\pm$$3.5$&  0.0&  0.0&$3.7$$\times$$1.6$&$ 51$&
$ 56.2$$\pm$$2.8$&  0.0&  0.0&$1.2$$\times$$0.6$&$ 90$
\\
&&&
E&
&&&&&
&&&&&
$  9.1$$\pm$$0.5$& -2.6&  2.1&$            <0.2$&$  0$
\\
&&&
E&
&&&&&
$  4.7$$\pm$$0.3$&  3.2&  1.6&$3.5$$\times$$1.5$&$ 32$&
&&&&
\\
&&&
C&
$  4.7$$\pm$$0.3$&  7.0&  1.8&$               -$&$  0$&
$  5.7$$\pm$$0.3$&  8.6&  2.0&$2.6$$\times$$2.0$&$ 96$&
&&&&
\\
&&&
W&
$ 66.5$$\pm$$3.3$& 16.0&  4.8&$6.4$$\times$$4.3$&$112$&
$ 14.6$$\pm$$0.8$& 15.0&  3.3&$4.4$$\times$$2.3$&$ 58$&
&&&&
\\
&&&
W&
$ 52.9$$\pm$$2.7$& 18.6&  4.8&$4.2$$\times$$2.1$&$ 33$&
$ 41.4$$\pm$$2.1$& 18.0&  5.2&$4.2$$\times$$1.9$&$ 38$&
$  7.9$$\pm$$0.4$& 25.5&  8.3&$            <0.1$&$  0$
\\
&&&
W&
&&&&&
&&&&&
$  4.4$$\pm$$0.3$& 27.2&  7.3&$            <0.2$&$  0$
\\
&&&
W&
&&&&&
&&&&&
$  2.6$$\pm$$0.2$& 28.1&  6.5&$            <0.2$&$  0$
\\
1841+6715&1&CD&
S&
$127$$\pm$$ 6$&  0.0&  0.0&$1.0$$\times$$0.8$&$126$&
$123$$\pm$$ 6$&  0.0&  0.0&$0.8$$\times$$0.5$&$169$&
$ 67.3$$\pm$$3.4$&  0.0&  0.0&$0.5$$\times$$0.4$&$142$
\\
&&&
N&
$ 17.8$$\pm$$0.9$&  1.4&  5.2&$2.5$$\times$$1.9$&$132$&
$  5.2$$\pm$$0.3$&  1.7&  5.9&$1.0$$\times$$0.9$&$ 71$&
$  2.6$$\pm$$0.2$&  2.3&  6.9&$            <0.4$&$  0$
\\
1843+6305&2&CD&
S&
$ 50.1$$\pm$$2.5$&  0.0&  0.0&$1.4$$\times$$0.7$&$156$&
$ 26.7$$\pm$$1.3$&  0.0&  0.0&$1.1$$\times$$0.7$&$177$&
&&&&
\\
&&&
N&
$ 17.8$$\pm$$0.9$&  2.6&  8.9&$2.3$$\times$$0.6$&$ 59$&
$  7.0$$\pm$$0.4$&  2.6&  9.1&$1.6$$\times$$1.0$&$ 17$&
&&&&
\\
1942+7214&1&CJ&
N&
$176$$\pm$$ 8$&  0.0&  0.0&$1.4$$\times$$0.8$&$ 58$&
$173$$\pm$$ 8$&  0.0&  0.0&$0.5$$\times$$0.4$&$  3$&
$101$$\pm$$ 5$&  0.0&  0.0&$0.4$$\times$$0.3$&$ 14$
\\
&&&
C&
$  3.5$$\pm$$0.3$&  6.1&-11.0&$               -$&$ 36$&
$  1.5$$\pm$$0.2$&  4.4&-10.8&$0.8$$\times$$0.8$&$  0$&
$  2.5$$\pm$$0.2$&  4.0&-11.8&$2.4$$\times$$0.7$&$ 55$
\\
&&&
S&
$  2.0$$\pm$$0.2$&  6.5&-17.4&$               -$&$171$&
&&&&&
&&&&
\\
&&&
S&
$  3.5$$\pm$$0.3$& 11.7&-29.3&$               -$&$  6$&
&&&&&
&&&&
\\
1945+6024&2&$-$&
C&
&&&&&
$ 81.3$$\pm$$4.1$&  0.0&  0.0&$0.6$$\times$$0.3$&$105$&
$ 79.7$$\pm$$4.0$&  0.0&  0.0&$            <0.1$&$  0$
\\
1946+7048&1&CSO&
N&
$292$$\pm$$14$&  0.0&  0.0&$1.6$$\times$$0.9$&$ 61$&
$215$$\pm$$10$&  0.0&  0.0&$1.6$$\times$$1.0$&$ 86$&
$106$$\pm$$ 5$&  0.0&  0.0&$1.1$$\times$$0.6$&$ 96$
\\
&&&
N&
$196$$\pm$$ 9$& -3.4& -0.1&$7.9$$\times$$3.0$&$ 83$&
$ 53.0$$\pm$$2.7$& -5.9& -0.5&$3.0$$\times$$1.5$&$ 95$&
$ 16.2$$\pm$$0.8$& -2.3& -0.6&$1.8$$\times$$1.0$&$ 73$
\\
&&&
N2&
$150$$\pm$$ 7$&  1.2& -6.6&$3.2$$\times$$1.2$&$  9$&
$ 81.0$$\pm$$4.1$&  1.2& -7.0&$2.9$$\times$$0.6$&$ 13$&
$ 40.7$$\pm$$2.0$&  1.2& -7.1&$2.1$$\times$$0.6$&$ 11$
\\
&&&
C&
$ 66.0$$\pm$$3.3$&  5.4&-13.3&$4.2$$\times$$1.4$&$ 47$&
$ 33.6$$\pm$$1.7$&  5.6&-13.5&$1.3$$\times$$0.2$&$ 45$&
$ 17.8$$\pm$$0.9$&  5.5&-13.2&$            <1.0$&$  0$
\\
&&&
C&
&&&&&
$ 27.6$$\pm$$1.4$&  7.4&-15.2&$0.9$$\times$$0.3$&$ 47$&
$ 66.5$$\pm$$3.3$&  6.5&-14.5&$1.2$$\times$$0.4$&$ 44$
\\
&&&
S2&
$ 76.4$$\pm$$3.8$& 10.3&-20.4&$3.0$$\times$$1.2$&$ 14$&
$ 41.4$$\pm$$2.1$& 10.2&-20.2&$1.5$$\times$$0.5$&$ 12$&
$ 18.2$$\pm$$0.9$&  9.9&-19.9&$1.0$$\times$$0.5$&$ 29$
\\
&&&
S&
$ 89.8$$\pm$$4.5$& 13.6&-28.9&$5.0$$\times$$2.6$&$ 51$&
$ 19.9$$\pm$$1.0$& 13.1&-29.0&$3.0$$\times$$1.1$&$ 44$&
&&&&
\\
1954+6146&2&CJ&
S&
&&&&&
$146$$\pm$$ 7$&  0.0&  0.0&$0.5$$\times$$0.2$&$153$&
$ 99.6$$\pm$$5.0$&  0.0&  0.0&$0.4$$\times$$0.1$&$141$
\\
&&&
N&
&&&&&
$  5.7$$\pm$$0.3$&  2.2&  3.8&$2.4$$\times$$1.5$&$159$&
&&&&
\\
1958+6158&2&$-$&
C&
&&&&&
$134$$\pm$$ 6$&  0.0&  0.0&$0.8$$\times$$0.4$&$ 60$&
$ 93.3$$\pm$$4.7$&  0.0&  0.0&$0.7$$\times$$0.4$&$ 87$
\\
\end{tabular}
\end{table*}

\section{Discussion}

In radio bright samples, GPS quasars are found to have
 core-jet or complex structure, while GPS galaxies are found
to have larger sizes with jets and lobes on both sides of a
putative center of activity (Stanghellini et al. 1997). Although observations
at another frequency are needed to confirm their classification, allmost all
radio-bright GPS galaxies from Stanghellini et al (1997) can be 
classified as CSOs.  The morphological dichotomy of GPS
galaxies and quasars, and their very different redshift distributions 
make it likely that GPS galaxies and quasars are not related to each
other and just happen to have similar radio spectra. 
It has been speculated that GPS quasars are a 
subset of flat spectrum quasars in general (eg. Snellen et al. 1999a). 
In addition, if galaxies and quasars were to be unified by orientation, 
due to changes in its observed radio spectrum (Snellen et al. 1998c),
it is not expected that a GPS galaxy observed at a small viewing angle would 
be seen as a GPS quasar. 
 
Not all CSOs are GPS sources. The contribution of the (possibly
variable) flat spectrum core can be significant and outshine the
convex spectral shape produced by the mini-lobes. This can be due to a
small viewing angle towards the object, causing the Doppler boosted
core and fast moving jet, which feeds the approaching mini-lobe, to be
important (Snellen et al. 1998c).  
An example of such a CSO, possibly observed at a small viewing
angle, is 1413+135 (Perlman et al. 1994).  In
addition, the jets feeding the mini-lobes can be significantly curved,
for example in 2352+495 by precession (Readhead et al. 1996).
This can cause parts of the jet to move at an angle close to the line
of sight, with significant Doppler boosting as a result.  In both
cases the large contrast between the approaching 
and receding parts of the radio source makes it also 
increasingly difficult to identify the object as a CSO.

 Figure \ref{class} shows the number of galaxies and quasars, in our faint 
GPS sample, classified as CJ, CSO, CX and those not possible to classify.  All
three objects classified as CSOs are optically identified with
galaxies. Although this is in agreement with the findings of
Stanghellini et al (1997) for the radio-bright sample, it should be
noted that for only 4 quasars was it possible to make a
classification. This is mainly due to the fact that the angular 
sizes of the quasars are significantly smaller than the angular sizes
of the galaxies.
 Six out of 18 classifiable GPS galaxies are found to
have CJ or CX structures, and 9 of the classifiable GPS galaxies are
found to have CD structures. We conclude that the strong 
 morphological dichotomy between
GPS galaxies and quasars found by Stanghellini (1997) in the bright
GPS sample, is not as strong in this faint sample. Note,
however, that the classification for the majority of the CJ and CD
sources is based on two components and their relative spectral indices
only. This makes their classification rather tentative. 
Firstly, a CD source could be erroneously
classified as a CJ source due to a difference in the observed age
between the approaching and receding lobe, causing a difference in
observed radio spectrum of the two lobes.  For a separation velocity
of 0.4c, as observed for radio bright GPS galaxies (Owsianik and
Conway 1998; Owsianik, Conway and Polatidis 1998), such an age difference can 
be as large as 30\%.
Secondly, differences in the local environments of the two lobes can
also influence the spectra of the two lobes, resulting in an
erroneous classification as core-jet.  For example, 
if only the two lobes had been visible, B1819+6707 (fig \ref{fig2})
could
have been mistaken for a core-jet source, since the spectral index of
the eastern lobe is flatter than that of the western lobe.

\section{Conclusions}

Multi-frequency VLBI observations have been presented of a faint
sample of GPS sources. All 47 sources in the sample were successfully
observed at 5 GHz, 26 sources were observed at 15 GHz, and 20 sources
were observed at 1.6 GHz. In this way 94\% of the sources have been
mapped above and below their spectral peak. The spectral
decomposition allowed us to classify 3 GPS galaxies as compact
symmetric objects (CSO), 1 galaxy and 1 quasar as complex (CX)
sources, 2 quasars and 5 galaxies as core-jet (CJ) sources, and 9
galaxies and 2 quasars as compact doubles (CD). 
Twenty-five of the sources could not be classified, 20 because they 
were too compact. The strong
morphological dichotomy of GPS galaxies and quasars found by
Stanghellini et al. (1997) in their radio bright GPS sample is not so
clear in this sample. However, many of the sources classified as CD and CJ 
have a two-component structure, making their classification only tentative.

\section*{Acknowledgements}

The authors are greatful to the staff of the EVN and VLBA for support of the 
observing projects. 
The VLBA is an instrument of the National Radio Astronomy Observatory,
which is operated by Associated Universities,
Inc. under a Cooperative Agreement with the National Science Foundation.
This research was supported by the European Commission, 
TMR Access to Large-scale Facilities programme under contract No. 
ERBFMGECT950012, and TMR Programme, Research Network Contract 
ERBFMRXCT96-0034 ``CERES''.

{}
\begin{figure*}
\centerline{\psfig{figure=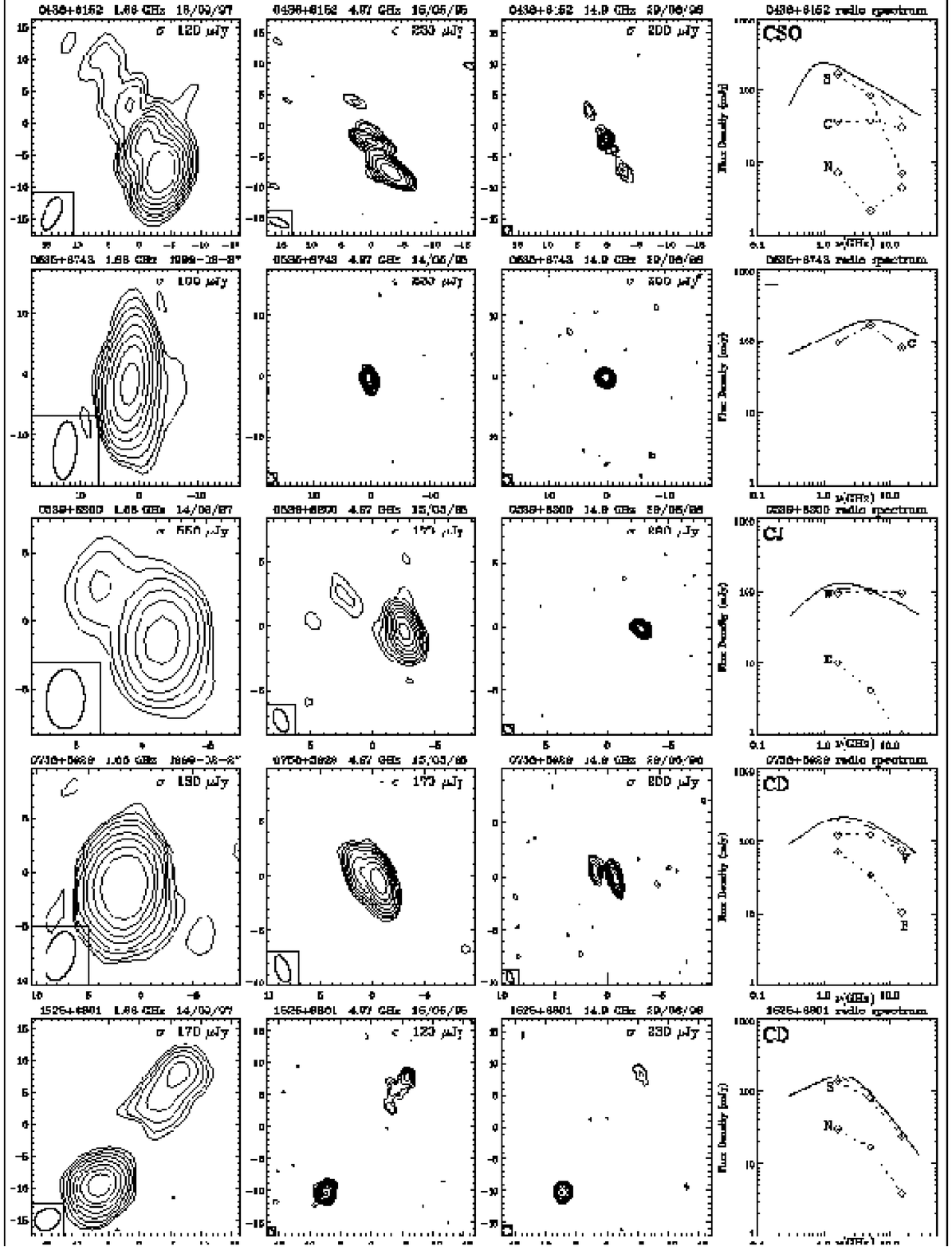,width=16cm}}
\caption{\label{fig3} The VLBI maps and spectral decomposition for sources observed at 3 frequencies. The noise level, $\sigma$ is given in the top right 
corner of each map. The contour levels are at $\sigma \times (-3, 3, 6, 12, 24, 48...)$. The solid line in the spectrum indicates the best fit to 
the overall radio spectrum, as derived in Snellen et al. (1998a).}
\end{figure*}
\addtocounter{figure}{-1}
\begin{figure*}
\centerline{\psfig{figure=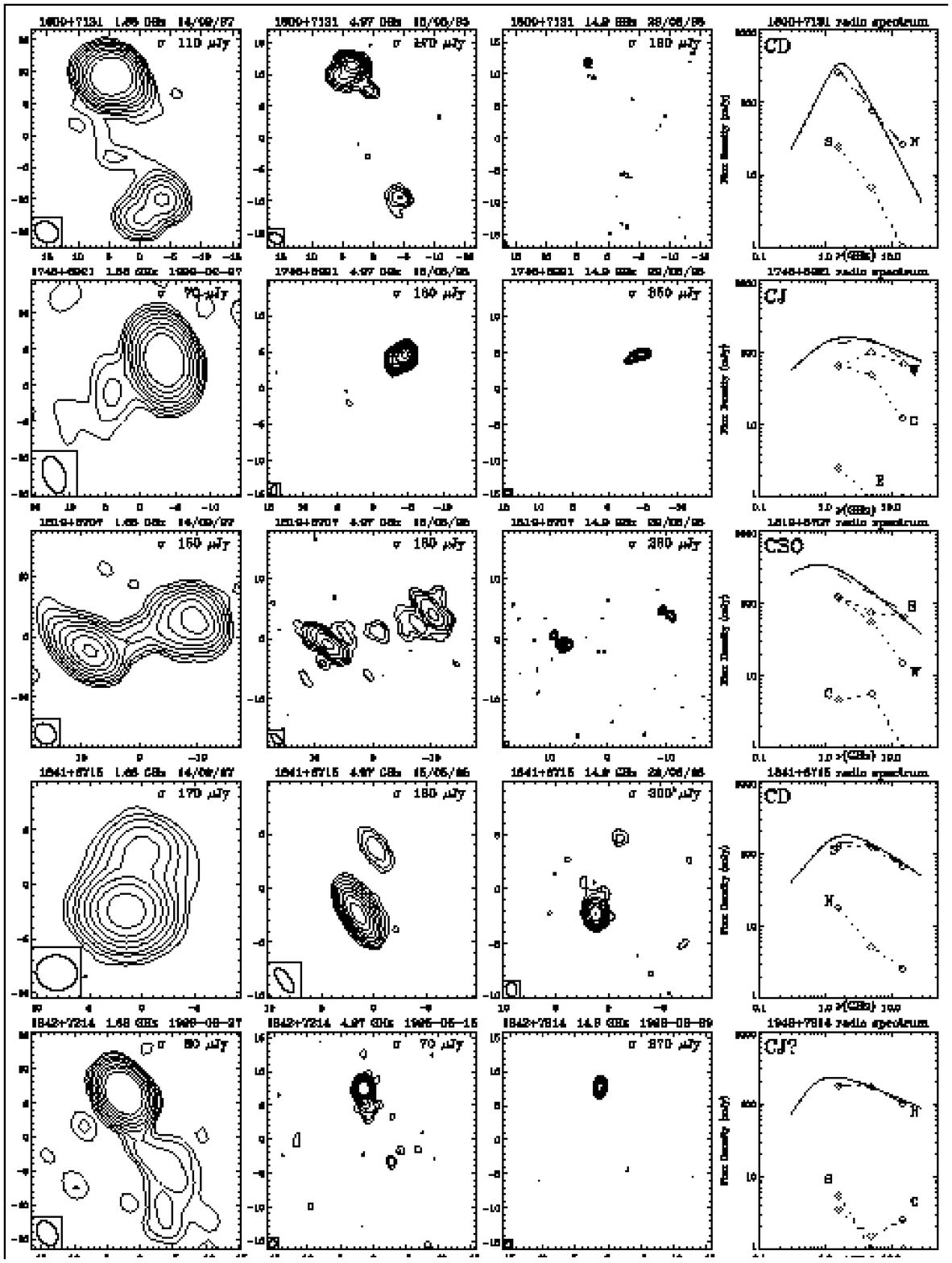,width=16cm}}
\caption{\label{fig3} Continued...}
\end{figure*}
\addtocounter{figure}{-1}
\begin{figure*}
\centerline{\psfig{figure=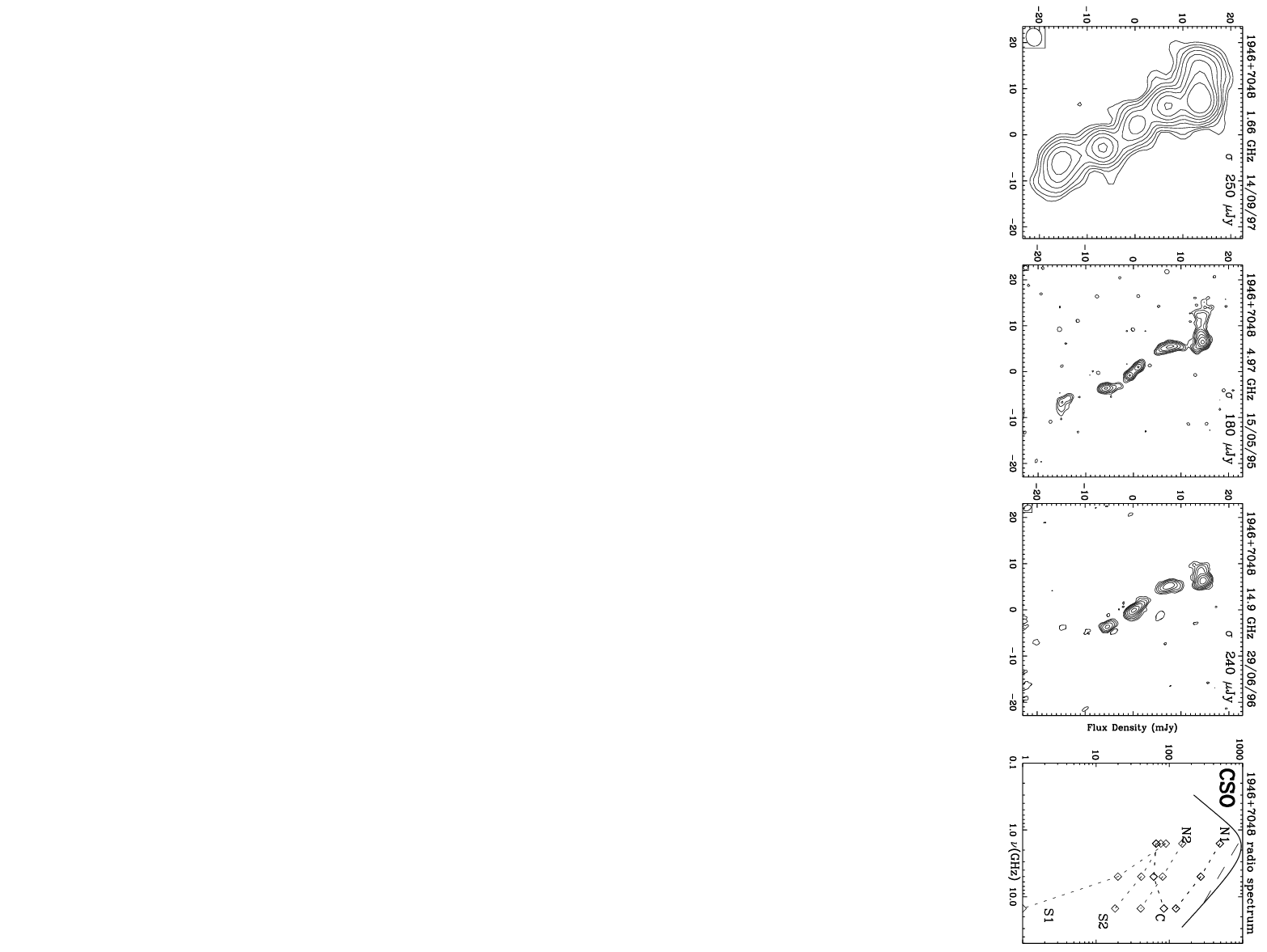,width=16cm,angle=90}}
\caption{\label{fig3} Continued...}
\end{figure*}

\begin{figure*}
\psfig{figure=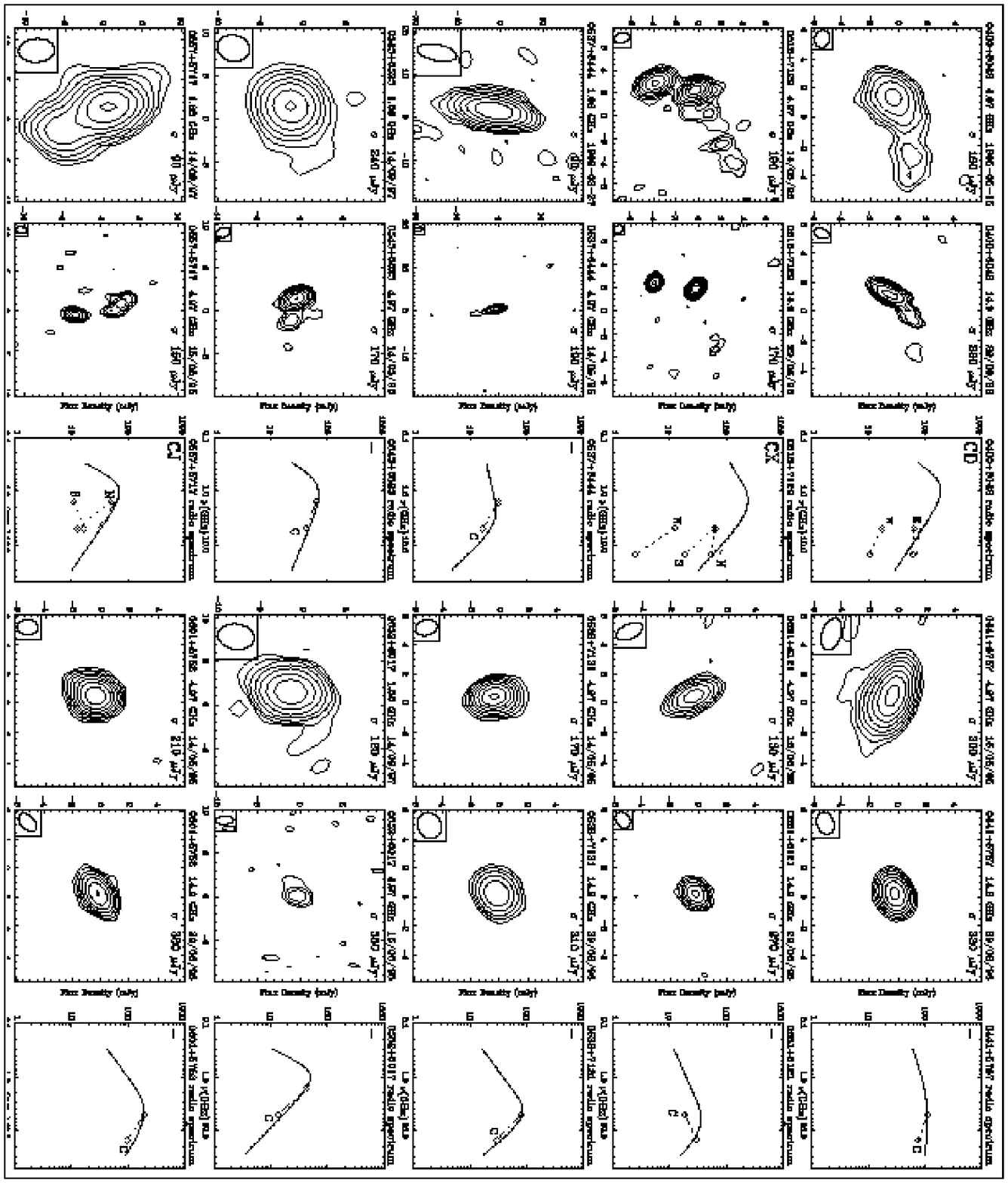,width=17cm,angle=180}
\caption{\label{fig2} The VLBI maps and spectral decomposition for sources observed at 2 frequencies. See table 2 for the beam sizes.
The noise level, $\sigma$ is given in the top right 
corner of each map. The contour levels are at $\sigma \times (-3, 3, 6, 12, 24, 48...)$. The solid line in the spectrum indicates the best fit to 
the overall radio spectrum, as derived in Snellen et al. (1998a).}
\end{figure*}
\addtocounter{figure}{-1}
\begin{figure*}
\psfig{figure=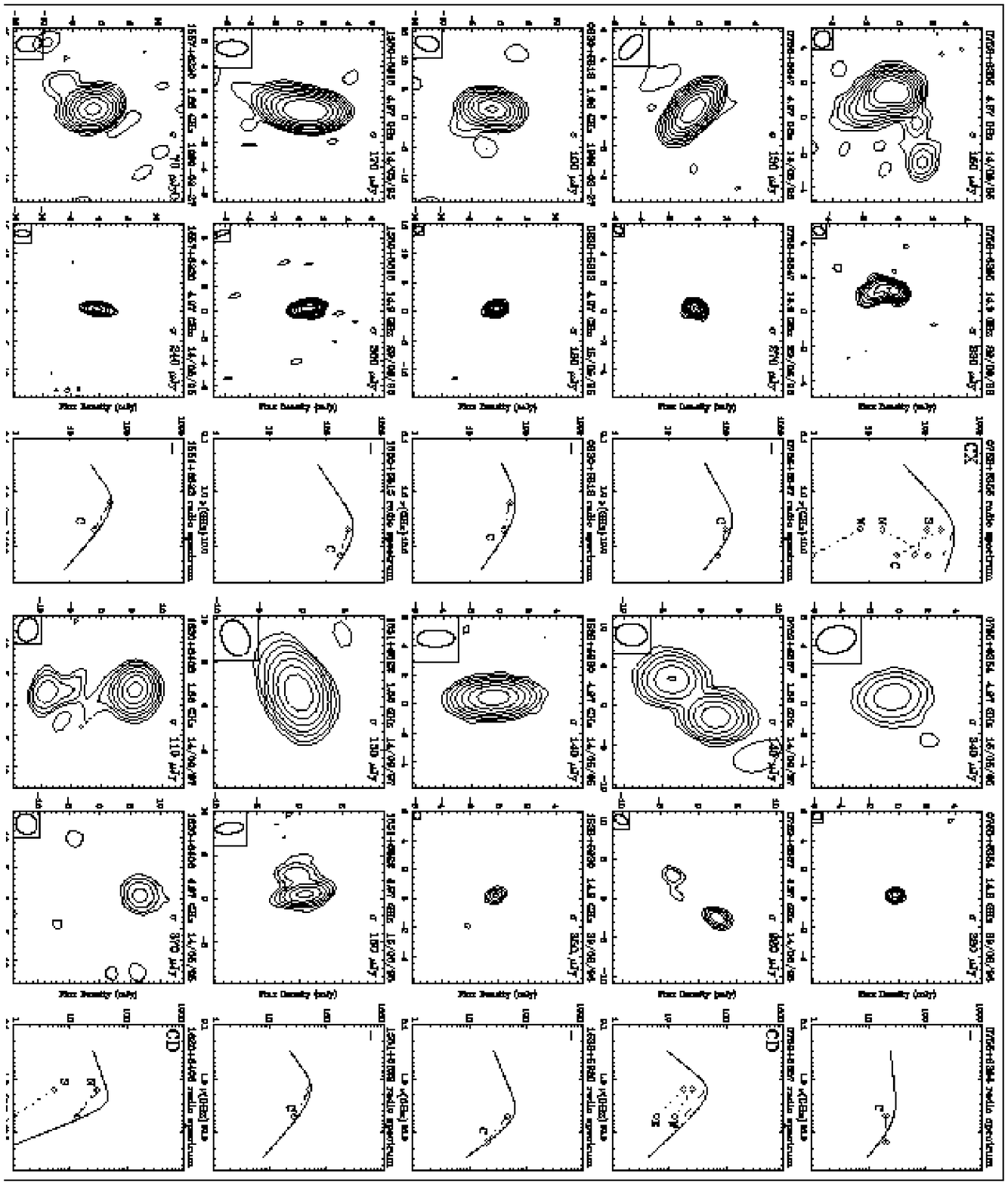,width=17cm,angle=180}
\caption{Continued...}
\end{figure*}
\addtocounter{figure}{-1}
\begin{figure*}
\psfig{figure=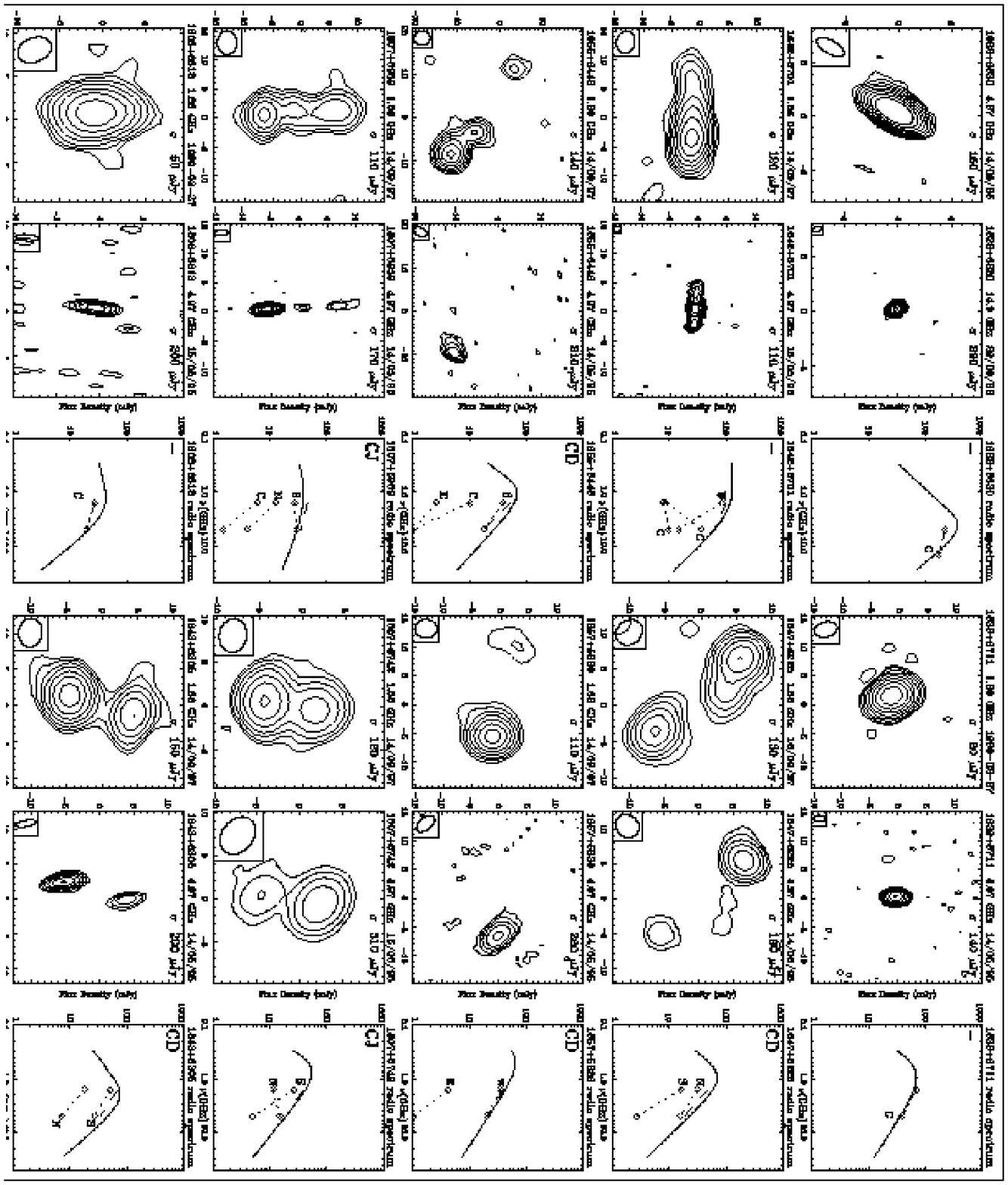,width=17cm,angle=180}
\caption{Continued...}
\end{figure*}
\addtocounter{figure}{-1}
\begin{figure*}
\centerline{\psfig{figure=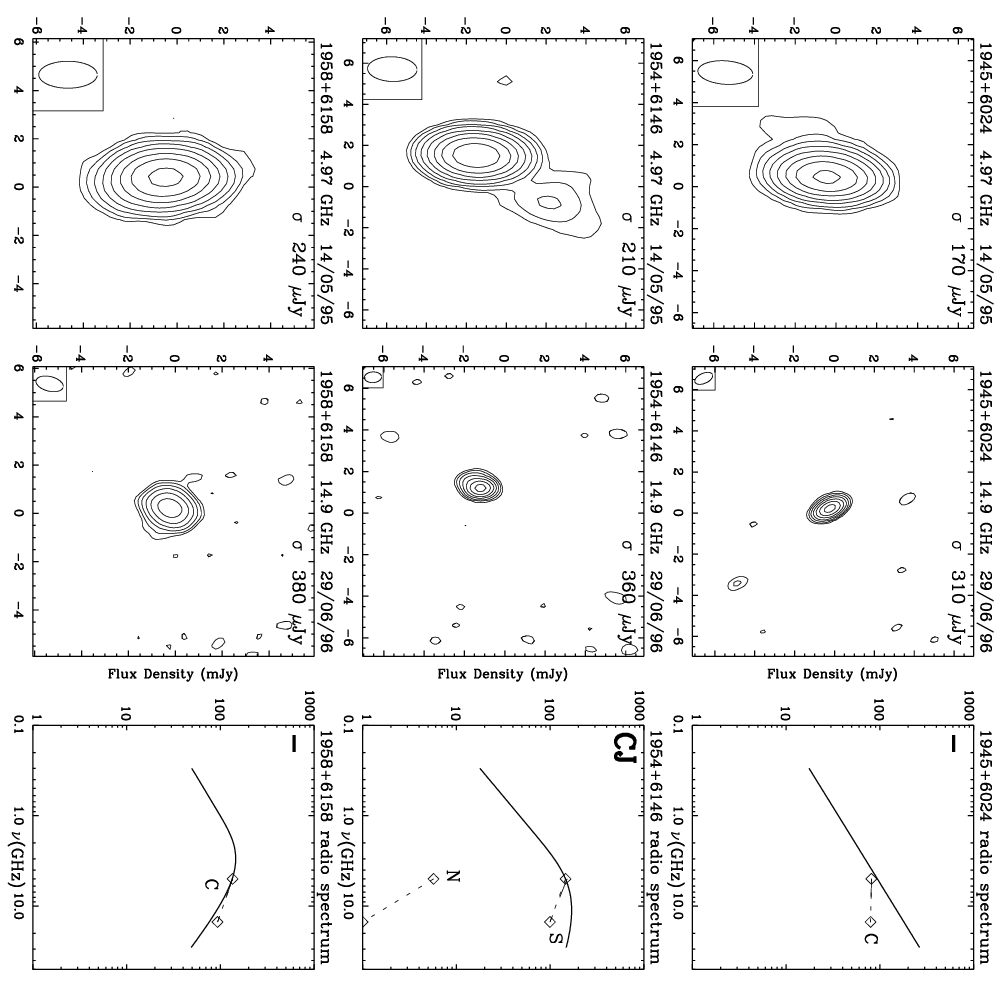,width=10cm,angle=90}}
\caption{Continued...}
\end{figure*}
\begin{figure*}
\centerline{\psfig{figure=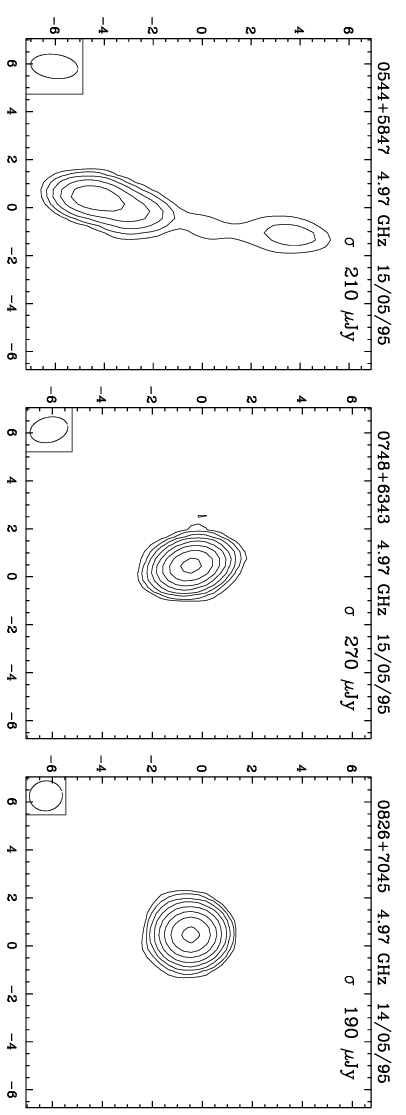,width=14cm,angle=90}}
\caption{\label{fig1} The VLBI maps for the sources observed only at 5 GHz.
The noise level, $\sigma$ is given in the top right 
corner of each map. The contour levels are at $\sigma \times (-3, 3, 6, 12, 24, 48...)$.}
\end{figure*}

\begin{figure}
\psfig{figure=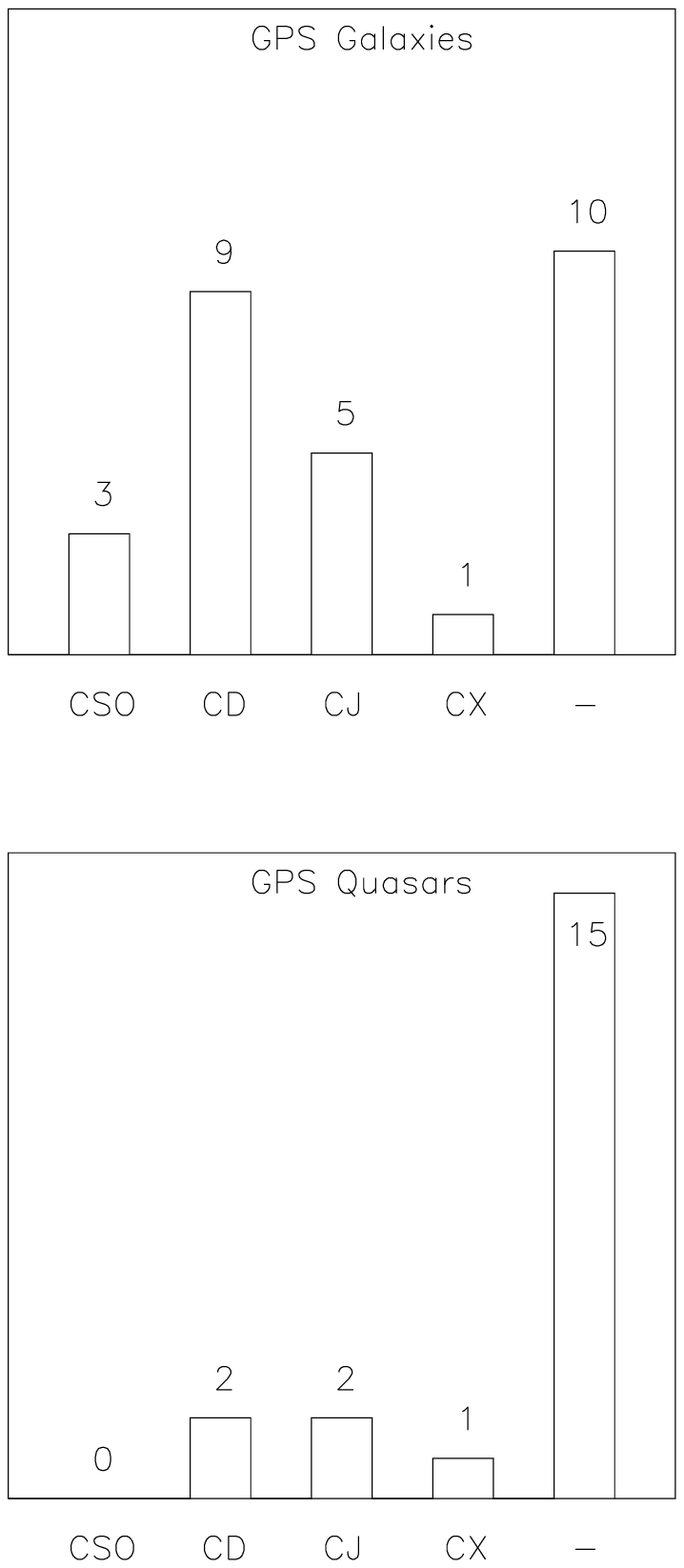,width=7cm}
\caption{\label{class}. The number of galaxies and quasars classified as
compact symmetric objects (CS0), Compact Doubles (CD), Core-jets (CJ), 
Complex sources (CX), and sources for which no classification was possible.}
\end{figure}
\end{document}